\documentclass{aa}

\usepackage{amsmath}
\usepackage{graphicx}
\usepackage[utf8]{inputenc}
\usepackage[usenames]{color}
\usepackage{txfonts}
\graphicspath{{Graficos/}}

\begin{document}

   \title{Near-Earth asteroids in Main Belt-crossing orbits}
   \author{P. S. Zain\inst{1,2}\thanks{pzain@fcaglp.unlp.edu.ar},
          R. P. Di Sisto\inst{1,2} \and R. Gil-Hutton\inst{3}}

   \offprints{Patricio Zain}
  \institute{Instituto de Astrofísica de La Plata, CCT La Plata-CONICET-UNLP. Paseo del Bosque S/N (1900), La Plata, Argentina.                             
   \and Facultad de Ciencias Astronómicas y Geofísicas, Universidad Nacional de La Plata. Paseo del Bosque S/N (1900), La Plata, Argentina.
    \and Grupo de Ciencias Planetarias, Dpto. de Geofísica y Astronomía, FCEFyN, UNSJ - CONICET, Av. J. I. de la Roza 590 oeste, J5402DCS Rivadavia, San Juan, Argentina.
   }

   \date{Received / Accepted}


\abstract
{}
{We study the dynamical and collisional evolution of Near-Earth asteroids (NEAs) in Main Belt–crossing orbits (NEACs).}
{We select NEACs with $H<18$ and integrate their orbits for $10^7$~yr with $N$-body simulations. Objects are grouped by initial semi-major axis (G1: $a<2.06$~au; G2: $2.06<a<2.5$~au; G3: $a>2.5$~au). We compute the fraction of each orbit spent within the main belt (MB), dynamical occupancy maps in the $(a,e)$ plane, and median lifetimes. Using collisional evolution, we obtain size-dependent timescales, the change in the NEA size-frequency distribution (SFD) over 1 Myr, and impactor and crater SFDs on 150 m to 1 km targets, representative of NEAs visited by space missions.
}
{Median dynamical lifetimes decrease with increasing $a$: $\sim1.3\times10^7$ yr (G1), $\sim2.1\times10^6$ yr (G2), and $\sim0.9\times10^6$ yr (G3). NEACs in G2–G3 maintain nearly constant MB residence fractions with short intervals of full containment, while G1 exhibits stronger 0–0.8 oscillations (median $\sim0.55$ for $\sim10^6$ yr). DART-analog impacts occur on $\sim10^5$ yr timescales for targets $\lesssim300$ m (rising to $\sim10^6$ yr for larger bodies), whereas catastrophic collisions are negligible within NEAC lifetimes. Over 1 Myr, collisional erosion reduces the meter-size NEA population by only $0.1$–$1.4$\% depending on $Q_D^*$. Comparison with the observed crater SFDs on Bennu, Didymos, and Ryugu indicates target strengths of $Y\approx100$ Pa for Bennu, young effective surface ages for Didymos, and short crater-retention times of order $10^{4}$--$10^{5}$ yr for craters with diameters $<100$ m on Ryugu, consistent with rapid resurfacing.}
{NEACs spend a substantial fraction of their lifetimes inside the MB and undergo frequent small-scale impacts, yet collisions weakly modifies the global NEA SFD on Myr timescales. Our combined dynamical–collisional framework constrains NEAC lifetimes, orbital pathways, collisional timescales, and surface processing.}

\keywords{minor planets, asteroids: general -- methods: numerical -- methods: statistical}
\authorrunning{P. S. Zain et al.}
\titlerunning{}

\maketitle
\section{Introduction}

\textit{Near-Earth asteroids} (NEAs) are a population of asteroids whose orbits have a perihelion distance $q$ of less than 1.3 au. The proximity of NEAs implies that their orbits can intersect Earth's orbit, which could eventually lead to an impact with our planet.

The source of NEAs is the Main Belt (MB), located between the orbits of Mars and Jupiter, at distances ranging from $\approx 2$ to $\approx 4$ au. Due to the combined effect of the Yarkovsky effect and orbital resonances with the giant planets, these asteroids experienced increases in their eccentricities that led them to migrate into the inner Solar System, where they frequently undergo close encounters with the rocky planets \citep{Granvik2017,Granvik2018}. NEAs have lifetimes of up to a few million years, and their typical final destinations are impacting the Sun, being ejected hyperbolically, or colliding with a terrestrial planet \citep{Bottke2001,Granvik2018}.

Recently, the \textit{Double Asteroid Redirection Test} (DART) mission was carried out to test the orbital deflection technique through a high-velocity impact. The target of the mission was a binary NEA system, whose primary component is Didymos, a 750 m asteroid, and whose secondary component and impact target is Dimorphos, a 150 m asteroid. The DART mission proved to be highly effective, as it reduced Dimorphos' orbital period around Didymos by approximately 30 minutes \citep{thomas2023orbital} and increased its momentum by a factor of between 2 and 5 times the incident momentum of the impact \citep{cheng2023momentum}.

Within the MB, mutual collisions between asteroids represent an important part of their evolution \citep{Bottke2015}. Depending on the impact energy, these collisions can be cratering or catastrophic. Collisional activity among NEAs is much less intense since their collisional lifetimes, particularly the catastrophic ones, are much longer than their dynamical lifetimes. However, Didymos is an asteroid that spends one-third of its orbit within the MB. In this context, \cite{campo2024recent} calculated the timescale on which Didymos and Dimorphos could be impacted by MB asteroids. 

In this work, we extend the study conducted by \cite{campo2024recent} to the entire population of NEAs that cross the MB (hereafter referred to as NEACs). We achieve this through $N$-body simulations and collisional evolution calculations.

This paper is organized as follows. In Section 2, we analyze the dynamical evolution of NEACs through $N$-body simulations. In Section 3, we present a collisional study of NEACs, including calculations of collisional timescales and the evolution of the NEA SFD distribution. Section 4 presents an estimation of the impactor flux and cratering rates over 1 Myr on representative targets, such as Bennu and Ryugu. Finally, our main conclusions are summarized in Section 5.

\begin{figure}
    \centering
    \includegraphics[width=9cm]{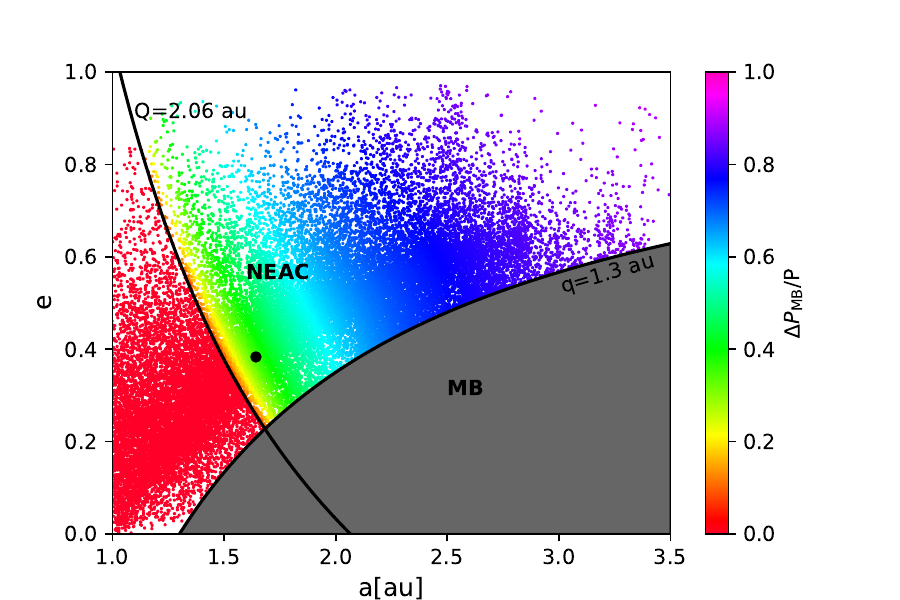}
    \caption{Distribution of semi-major axes $a$ and eccentricities $e$ of NEAs. The color bar indicates the fraction of time that NEAs spend within the MB relative to their orbital period. The curve $q=1.3$ au marks the boundary between NEAs and the MB (gray region), while NEACs correspond to the region to the right of the curve $Q=2.06$ au. The black dot indicates the $a$ and $e$ values of Didymos.}
    \label{fig:NEACae}
\end{figure}

\section{Incursions in the MB}

\begin{figure}
    \centering
    \includegraphics[width=8.5cm]{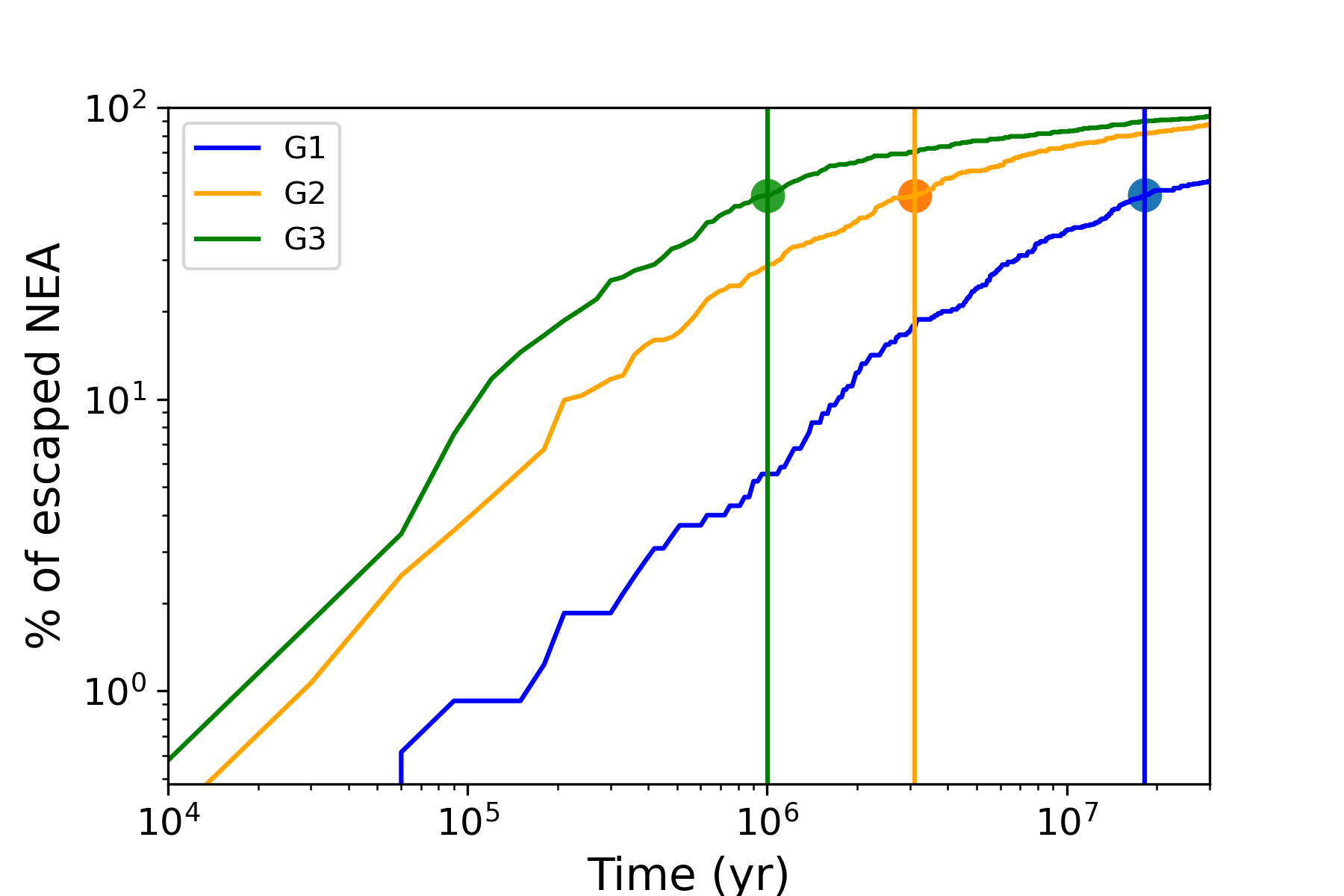}
    \caption{Cumulative percentage of escaped NEACs as a function of time for Groups 1–3. Escapes are defined as reaching a heliocentric distance $>100$\,au or colliding with the Sun or a planet. Solid lines show the evolution for G1 (blue), G2 (orange), and G3 (green), while vertical dashed lines and markers indicate the median lifetime.}
    \label{fig:escape}
\end{figure}

\begin{figure*}
    \centering
    \includegraphics[width=18cm]{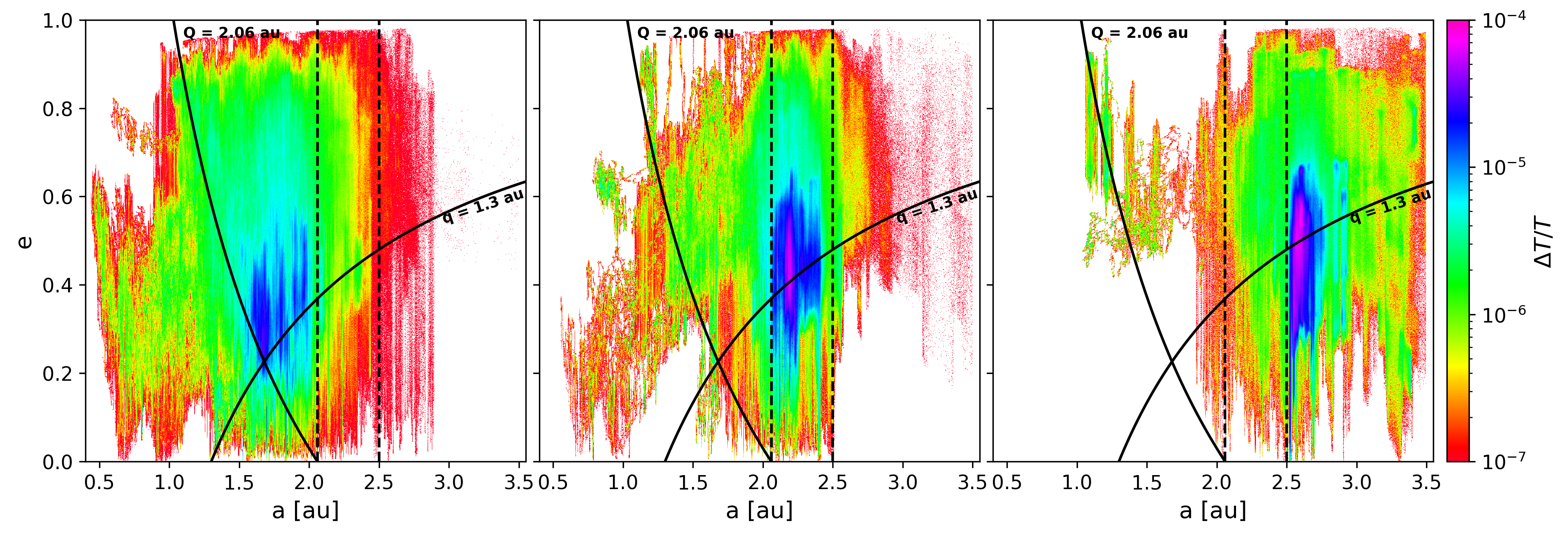}
    \caption{Dynamical occupancy maps in the $a$–$e$ plane for G1 (left panel), G2 (middle panel) and G3 (right panel). Colors indicate the fraction of total integration time spent in each of the 1000 grid cells. Solid black curves mark perihelion $q=1.3$\,au and aphelion $Q=2.06$\,au; dashed vertical lines denote the group boundaries at $a=2.06$\,au and $a=2.5$\,au.}

    \label{fig:ae}
\end{figure*}

\begin{figure*}
    \centering
    \includegraphics[width=6cm]{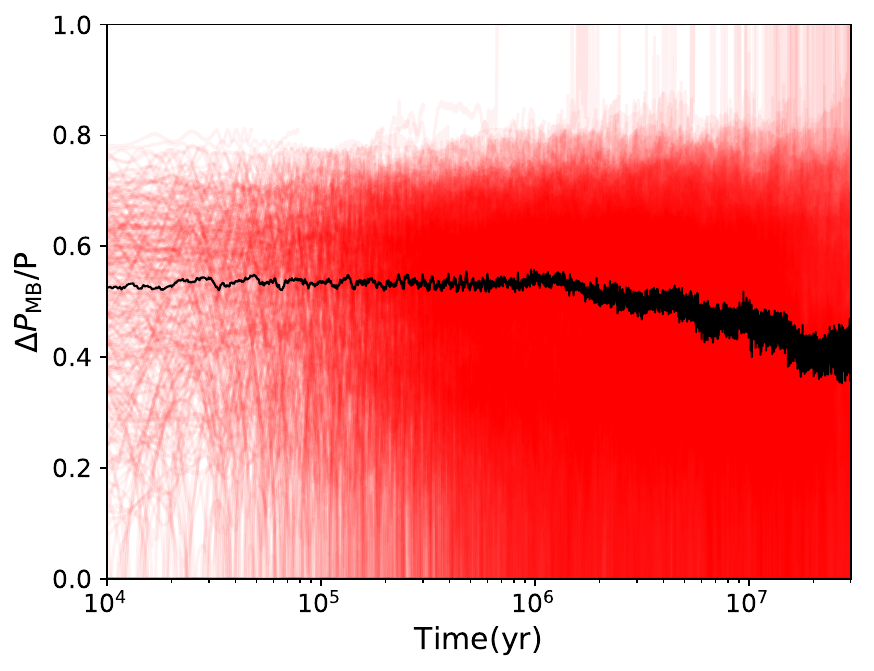}
    \includegraphics[width=6cm]{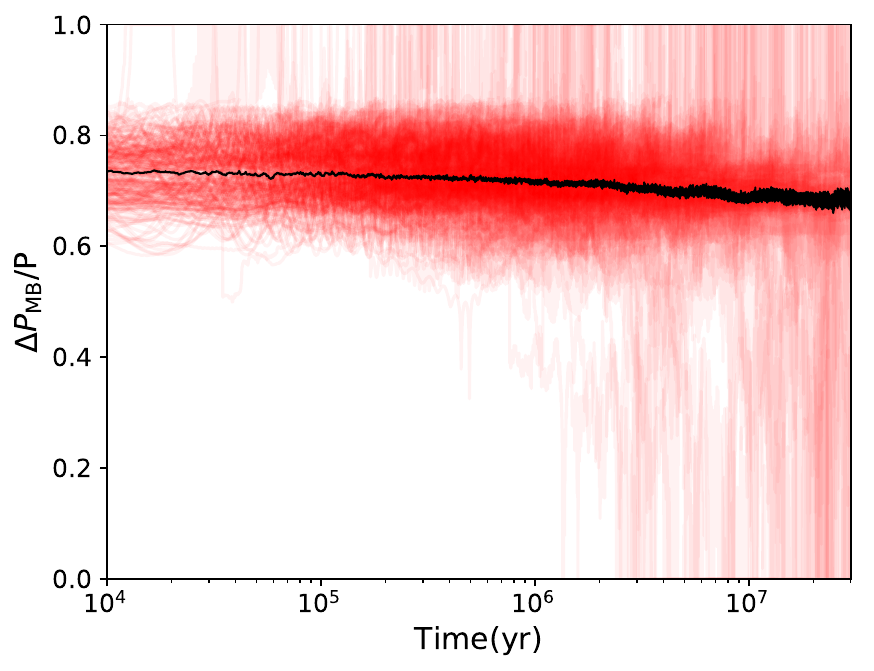}
    \includegraphics[width=6cm]{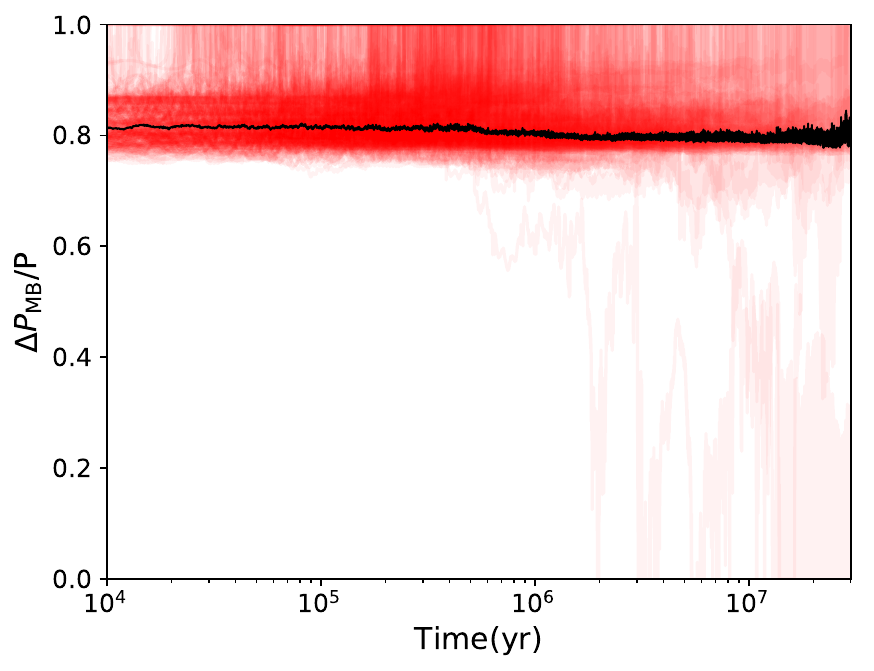}
    \caption{Evolution of the fraction of time spent in the MB relative to the orbital period, $\Delta P_\mathrm{MB}/P$, for each NEAC (red) and the median at each time step (black) for G1 (left panel), G2 (middle panel), and G3 (right panel).}
    \label{fig:dT}
\end{figure*}

In this section, we analyse the dynamical evolution of NEAC orbits. Fig. \ref{fig:NEACae} shows the semi-major axes $a$ and eccentricities $e$ of all NEAs, obtained from the \textit{Minor Planet Center}\footnote{https://www.minorplanetcenter.net/data}. We formally define NEACs as those whose aphelion is at a distance $Q > 2.06$ au, which corresponds to the inner boundary of the MB.  

For each NEAC, we compute the time its orbit ventures into the MB using Kepler’s equation:

\begin{equation}
    E - e\sin{E} = \frac{n}{2} (P - \Delta P_\mathrm{MB}),
\end{equation}

where $E$ is the eccentric anomaly when the orbit crosses the MB (i.e $r = 2.06$) , $n$ is the mean orbital motion, $P$ is the orbital period of the asteroid and $\Delta P_\mathrm{MB}$ is the time inside the MB. The fraction $\Delta P_\mathrm{MB}/P$ is represented by the colorbar in Fig. \ref{fig:NEACae}. We obtain that NEACs spend a mean fraction of 0.6 of their orbital periods within the MB. However, these fractions are calculated using the two-body problem with their current orbital elements. To determine how NEAC orbits and their incursions into the MB evolve over their lifetimes, we perform $N$-body simulations using the \texttt{MERCURY} code \citep{chambers1999hybrid}.

In these simulations, we consider NEACs with absolute magnitude $H < 18$, where the population is believed to be complete \citep{Tricarico2017}, and we include the eight planets of the Solar System as massive bodies. We classify NEACs into three groups according to the value of $a$: Group 1 (G1) includes those with $a < 2.06$ au, Group 2 (G2) those with $2.06 < a < 2.5$ au, and Group 3 (G3) those with $a > 2.5$ au. The number of NEACs in each group is 325, 282 and 290, respectively. 

The chosen boundaries, $a = 2.06$ and $a = 2.5$ au, correspond to the inner edge of the MB and the location of the 3:1 mean-motion resonance with Jupiter, which separates the inner and middle MB regions \citep{deElia2007,Zain2020} and constitutes one of the main sources of NEAs \citep{Bottke2001,Granvik2018}. We use the hybrid integrator of \texttt{MERCURY} and evolve each group of NEACs as massless particles for a maximum time of $3 \times 10^7$ years.

Fig. \ref{fig:escape} shows the percentage of NEACs that escape the simulation as a function of time for each of the three groups defined previously. Escapes are identified when an NEAC either reaches a heliocentric distance greater than 100 au, collides with the Sun or a planet. The vertical lines and solid markers indicate the median dynamical lifetime for each group. A clear dependence on the initial semi-major axis is observed: NEACs from G3 are the most dynamically short-lived, with a median lifetime of $\sim 1 \times 10^6$ years, followed by G2 with $\sim 3 \times 10^6$ years. In contrast, NEACs from G1 show significantly longer dynamical lifetimes, with a median value close to $2 \times 10^7$ years. The shorter lifetimes of NEACs in G2 and G3 can be attributed to stronger gravitational perturbations from the giant planets, particularly Jupiter. 

Fig. \ref{fig:ae} displays the dynamical occupancy maps for all three NEAC groups in the $a$--$e$ plane over the full integration. The left panel shows that the G1 objects occupy a broadly dispersed region extending from $a \approx 0.5$~au up to 3 au and spanning the full eccentricity range. Their orbits are initially belt-crossing, and along their evolution individual G1 objects can reach configurations entirely contained within the MB ($q > 1.3$ au) or entirely interior to the belt, i.e. outside the MB region ($Q < 2.06$ au). In contrast, G2 NEACs (middle panel of Fig.~\ref{fig:ae}) predominantly act as belt-crossers, yet can also transiently reach orbits entirely contained within the MB. Survivors beyond 1~Myr cluster around a stable region near $a \approx 2.2$~au and $q \approx 1.3$~au, where they oscillate between belt-crossing and fully confined trajectories, a similar behaviour to that found for G3 in the right panel of Fig.~\ref{fig:ae}. This orbital picture is consistent with the evolution of the fraction of time spent in the MB relative to the orbital period, $\Delta P_{\mathrm{MB}}/P$, illustrated in Fig. \ref{fig:dT}. NEACs from groups G2 (middle panel) and G3 (right panel) maintain a roughly constant fraction of their orbit within the MB over their lifetimes, with median values of $\approx 0.75$ and $\approx 0.8$, respectively. The short vertical segments, where individual curves spike up to $\Delta P_{\mathrm{MB}}/P = 1$, highlight intervals in which bodies remain fully confined to the MB region for several consecutive time steps before reverting to belt-crossing orbits. In contrast, the evolution is more chaotic for NEACs from G1 (left panel), where the fraction oscillates between 0 (orbits entirely interior to the MB, outside the belt region) and values up to 0.8 (belt-crossing trajectories), with a median that remains at $\approx 0.55$ for approximately $10^{6}$~yr. Overall, NEACs in our sample can travel back and forth between the MB and the inner Solar System several times, as their orbits alternate between configurations interior to the MB, belt-crossing trajectories, and orbits entirely contained within the belt.

From this first part of the study, we can conclude that NEACs indeed spend a large portion of their time within the MB and, therefore, can be impacted by MB asteroids. This collisional activity will be analysed in the next section.

\section{Collisional activity}

\subsection{Collisional lifetimes}

\begin{figure}
    \centering
    \includegraphics[width=8.5cm]{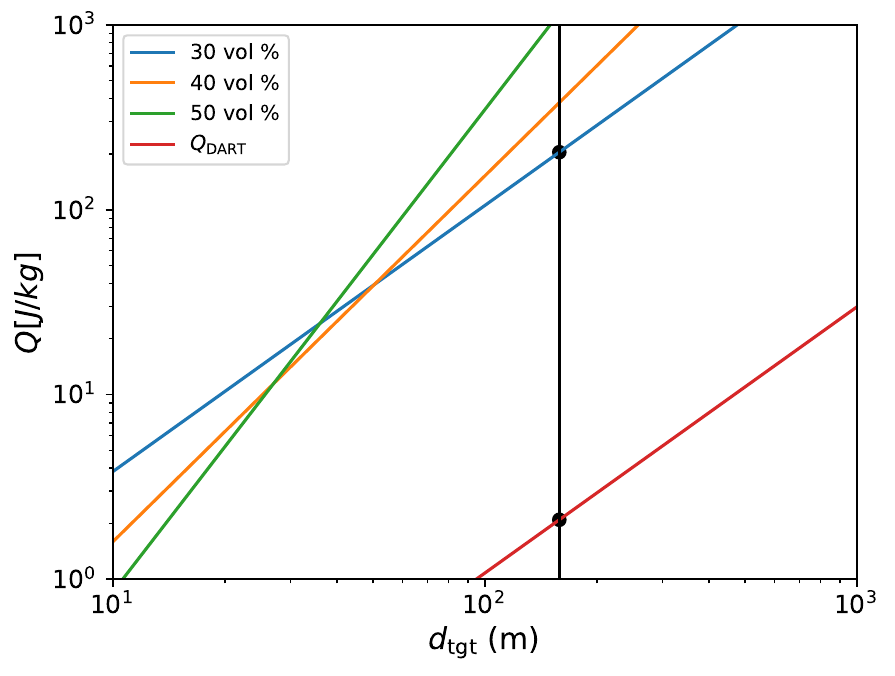}
    \caption{$Q_D^*$ as a function of target diameter $d_\mathrm{tgt}$ for rubble‐pile asteroids with different large‐boulder volume fractions \citep{raducan2024lessons}. $Q_\mathrm{DART}$ corresponds to the 30\% boulder‐volume case scaled to the DART impact energy.}

    \label{fig:QD}
\end{figure}

In this section, we analyse the collisional activity between targets that are NEACs and impactors that are MBAs. A relevant quantity for studying this activity is $Q_{\text{D}}^*$, defined as the energy per unit mass required to disperse half of the mass of a target. This quantity marks the energy threshold between cratering and catastrophic collisions. In this section, we use the formulation proposed by \cite{raducan2024lessons}, who calculated and calibrated $Q_{\text{D}}^*$ for \textit{rubble-piles} with different bulk densities and spherical large rock distributions, using data from the DART mission. These functions are presented in Fig. \ref{fig:QD} and show that the energy required to catastrophically break an asteroid varies significantly depending on the volume occupied by large rocks. It is important to note that this property of Dimorphos' internal structure is unknown until the arrival of the Hera mission, which will study the asteroid's state after the impact.  However, to establish a lower bound for catastrophic collisions, we consider in this work the value of $Q_{\text{D}}^*$ with a large rock bulk density of $30\%$ of the target's volume. With this value and the impact energy of the DART mission, $E_\mathrm{DART} = 1.094 \times 10^{10} \ \mathrm{J}$ \citep{daly2023successful}, we scale $Q_{\text{DART}}$, a function that approximates the specific energy required for an impact on an NEA of a given size to be analog to that of DART. The values of $Q_\mathrm{D}^*$ can be converted to the diameters of impactors using $d_\mathrm{imp}=\left(2Q_\mathrm{D}^*/v^2\right)^{1/3} d_\mathrm{tgt}$ \citep{Bottke2005a}, where $v$ is the impact velocity. We adopt $v = 7.5~\mathrm{km\,s^{-1}}$, as calculated for Didymos by \citet{campo2024recent}.

The impact frequency between a target of size $d_\mathrm{tgt}$ and an impactor of size $d_\mathrm{imp}$ is given by:

\begin{equation}
    f(d_{\rm tgt},d_{\rm imp}) = \frac{P}{4} \left(d_{\rm tgt}+d_{\rm imp}\right)^{2},
    \label{eq:Frec}
\end{equation}

where $P=1\times10^{-17} ~ \mathrm{km}^{-2}\text{yr}^{-1}$ \citep{campo2024recent} is the assumed intrinsic collision probability and the diameters are expressed in km.

From this, the characteristic timescale for a collision by an impactor with diameter $d_\mathrm{imp}$ onto a target with diameter $d_\mathrm{tgt}$ is given by:

\begin{equation}
\tau = \frac{1}{f(d_{\rm tgt},d_{\rm imp})\,N(d_{\rm imp})},
\end{equation}  
where $N(d_{\rm imp})$ is the SFD of MB asteroids from \citet{Bottke2020}.

We adopt a single representative value of $P$ equal to that derived by \citet{campo2024recent} for the Didymos system, a NEAC belonging to our G1 group. The intrinsic collision probability depends on the detailed orbital elements of each target, and for Didymos, \citet{campo2024recent} estimated $P = 1\times10^{-17}$ km$^{-2}$ yr$^{-1}$ with an uncertainty of about 10\%. Since the collision rates scale linearly with $P$ and the collisional lifetimes as $\tau \propto P^{-1}$, this uncertainty may translate into variations of only $\pm 10$\% in the derived timescales of this work.

\begin{figure}
    \centering
    \includegraphics[width=8cm]{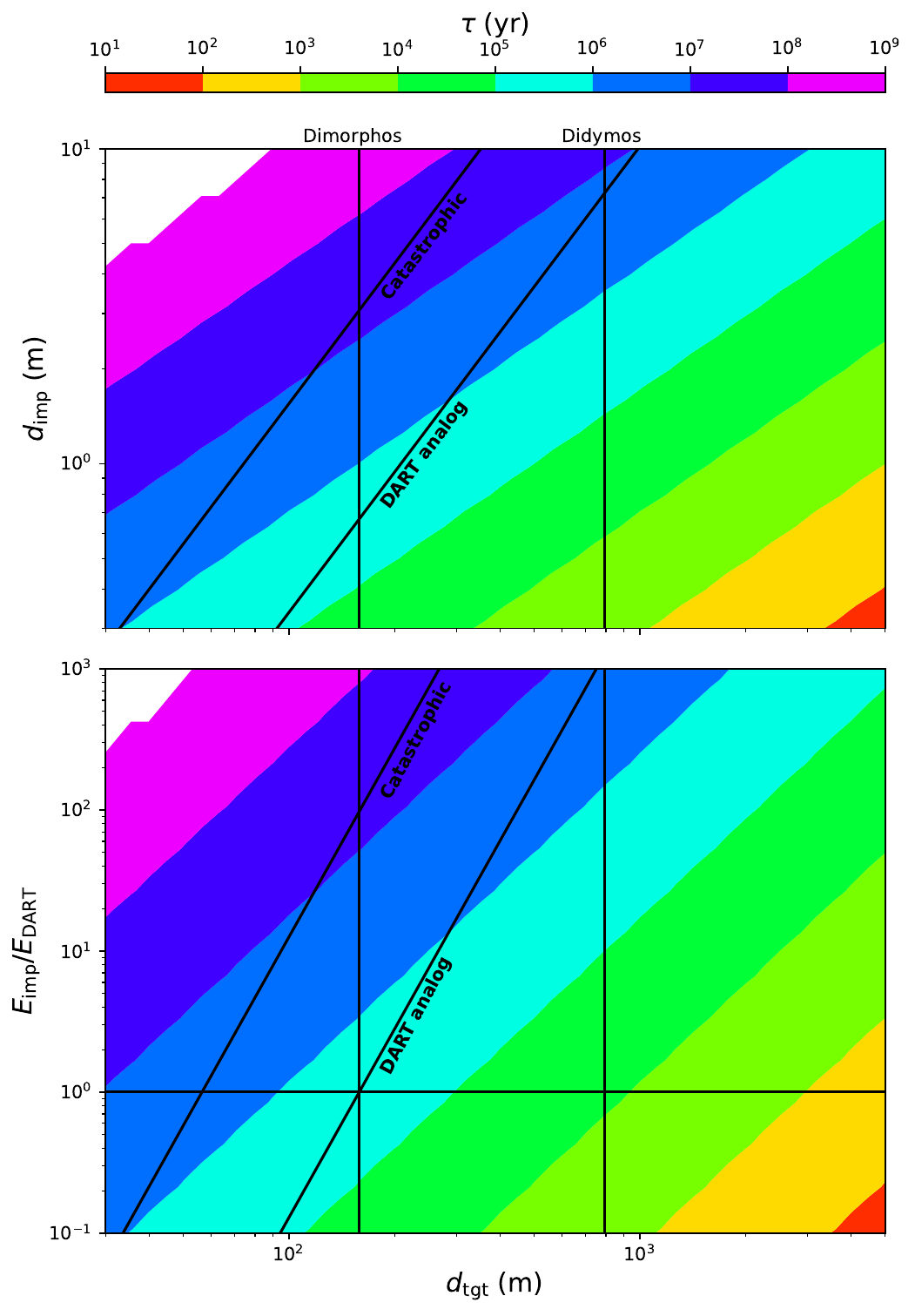}

    \caption{Collisional timescales $\tau$ (color bar, in years) for NEACs as a function of target diameter $d_{\rm tgt}$ and impactor diameter $d_{\rm imp}$ (top), and normalized impact energy $E_{\rm imp}/E_{\rm DART}$ (bottom). Solid black lines in both panels mark the energy thresholds for DART‐analog impacts and catastrophic disruption. Vertical lines indicate the diameters of Didymos (750 m) and Dimorphos (150 m). The horizontal line in the lower panel denotes $E_{\rm imp}/E_{\rm DART}=1$, where impact energies equal that of the DART mission.}

    \label{fig:Tiempos}
\end{figure}

Fig. \ref{fig:Tiempos} shows, through a colour bar, the orders of magnitude of the collisional timescales for different target-impactor pairs as a function of impactor diameter in the upper panel, and the impact energies relative to the DART mission energy in the lower panel. Since the dynamic lifetimes of NEAs reach at most a few million years \citep{Granvik2018}, events located below the light blue band, corresponding to $10^{5}$ years, can occur multiple times during this period. In particular, impacts with the exact DART energy (horizontal line in the lower panel) are frequent in NEACs of all sizes, occurring as often as every $10^{5}$ years in objects up to 300 m, with shorter timescales as the target diameter increases. In contrast, events located within the blue band, corresponding to $10^{6}$ years would be limited to a few cases depending on the exact dynamic lifetime of the NEACs. If we consider impacts with energy comparable to DART, these occur on timescales of $10^{5}$ years for NEACs up to 300 meters, while for larger sizes, the timescales increase to $10^{6}$ years.
Likewise, we can rule out the occurrence of catastrophic collisions in NEACs larger than 100 m, as the characteristic timescales for these events exceed $10^{7}$ years.

\subsection{Change in the NEA SFD}

Here, we estimate how the NEA\footnote{For simplicity, in this section we use the SFD of the entire NEA population instead of the NEACs defined in previous sections.} SFD evolves due to catastrophic collisions with the MB population over a period of 1 Myr. To do this, we adopt the NEA SFD derived from NEOMOD3 \citep{nesvorny2024neomod} and the MB SFD from \citet{Bottke2020}. The change in the differential NEA SFD over a single timestep $\Delta t$ is given by:

\begin{equation}
\Delta_\mathrm{NEA}(d_\mathrm{tgt},t) = - f(d_\mathrm{tgt},d_\mathrm{imp})N(d_\mathrm{tgt})N(d_\mathrm{imp}) \Delta t,
\end{equation}

where the right-hand side represents the number of collisions between targets and projectiles. Here, $N(d_\mathrm{tgt})$ and $N(d_\mathrm{imp})$ are the differential SFDs of the target and impactor populations, respectively. The negative sign reflects the fact that we only consider the removal of bodies due to catastrophic collisions, defined as those where the specific impact energy $Q$ exceeds the disruption threshold $Q_\mathrm{D}^*$—the energy required to disperse more than half the target's original mass. We do not take into account the fragment generation in collisions, which can increase the number of small bodies in the SFD. We perform simulations using four different prescriptions for the scaling law $Q_\mathrm{D}^*$,  the basaltic target law from \citet{BenzAsphaug}, prescriptions \#1 and \#8 from \citet{Bottke2020}, and the rubble-pile scaling law from \citet{raducan2024lessons} from Section 3.1.

\begin{figure}
    \centering
    \includegraphics[width=8cm]{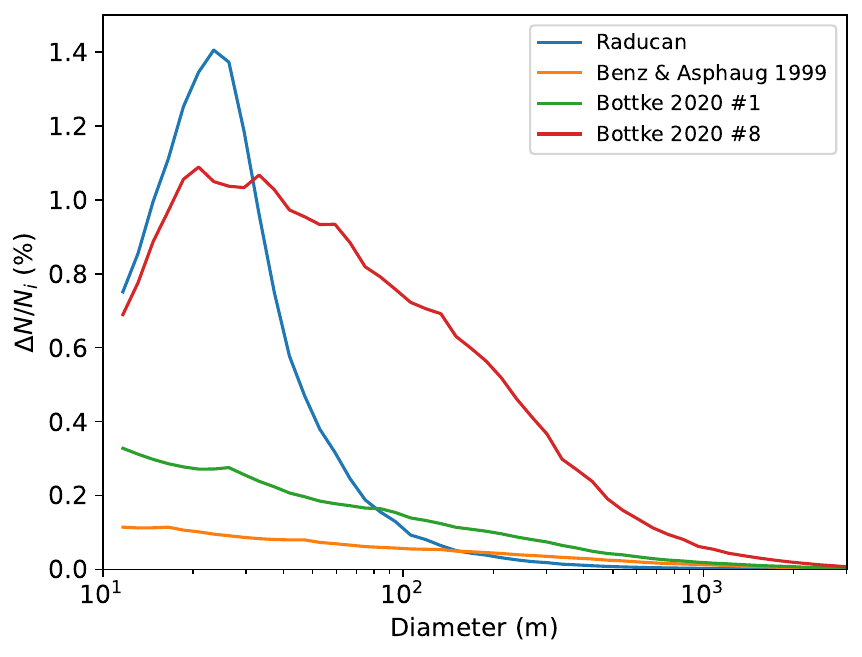}
    \caption{Change in the NEA SFD due to the collisional evolution between NEA targets and MB projectiles over 1 Myr, for different scaling laws considered for $Q_D^*$. }
    \label{fig:CambioSFD}
\end{figure}

Fig.~\ref{fig:CambioSFD} shows the change in the NEA SFD relative to the NEOMOD3 reference after 1 Myr of collisional evolution. Catastrophic collisions with MB projectiles produce only minor modifications to the NEA population, regardless of the scaling law adopted. In particular, stronger disruption thresholds such as those of \citet{BenzAsphaug} and model \#8 from \citet{Bottke2020} result in reductions of up to 0.1\% and 0.4\%, respectively. Conversely, weaker prescriptions like that of \citet{raducan2024lessons} and model \#1 from \citet{Bottke2020} lead to larger decreases of up to 1.4\% and 1.1\%. The most significant depletion is observed at intermediate diameters, roughly between 10 m and 300 m, and above $\sim$1~km the curves converge to negligible values.

\section{Impactors and cratering on specific asteroids} 

Recent missions to NEAs have delivered detailed measurements of their physical properties and high-precision global crater catalogs. Building on the NEAC–MB collisional activity quantified in the previous section, we estimate the number of impactors striking NEACs across a range of various target sizes representative of specific asteroids targeted by recent space missions, such as Dimorphos, Bennu, Didymos, and Ryugu. We calculate the number of impacts between a target of size $d_\mathrm{tgt}$ and impactors of size $d_\mathrm{imp}$ using Poisson statistics with mean $f(d_\mathrm{imp}, d_\mathrm{tgt}) N(d_\mathrm{imp}) \Delta t$, where $N(d_\mathrm{imp})$ is the differential SFD of the MB impactors from \cite{Bottke2020}. We adopt time-steps $\Delta t$ of 1000 years and simulate the collisional evolution over a total duration of 1 Myr. This process is repeated 1000 times for each target size to capture the statistical variability in the number of impacts.

\begin{figure}
    \centering
    \includegraphics[width=8cm]{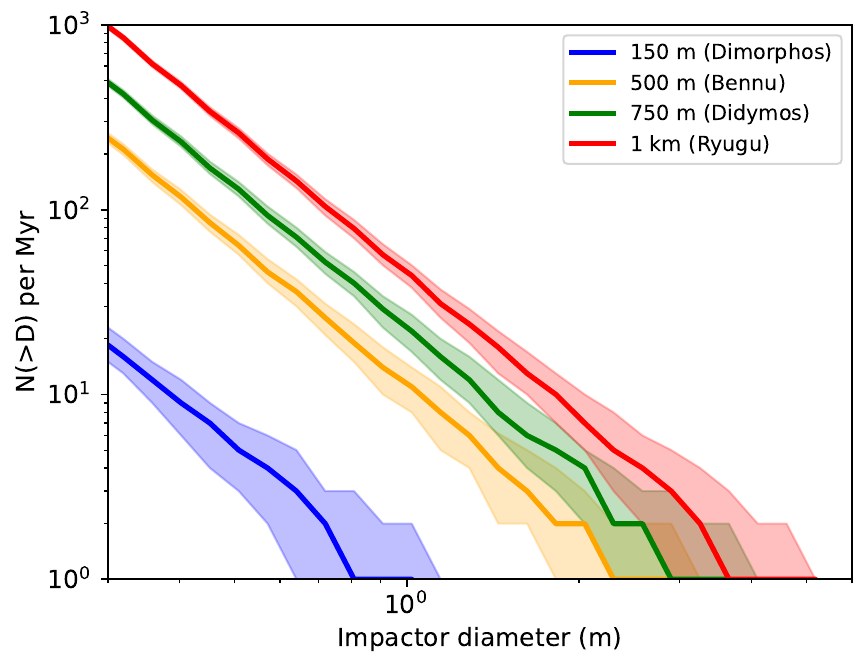}
    \caption{SFDs of impactors onto NEA targets of 150 m, 500 m, 750 m, and 1 km in diameter over 1 Myr. Solid lines indicate the median value in each size bin, while the shaded regions indicate the 16th–84th percentile range across 1000 simulations.}

    \label{fig:Impactores}
\end{figure}

Fig. \ref{fig:Impactores} shows the resulting SFDs of impactors for four representative NEA targets with diameters of 150 m, 500 m, 750 m, and 1 km. Solid curves represent the median number of impactors per diameter bin. The shaded regions enclose the 16th to 84th percentiles, which reflects the statistical dispersion across simulations. The resulting SFDs generally follow a power-law trend with slope $\approx-2.54$. The largest impactors in the median SFDs have diameters of 1.00 m, 2.89 m, 3.65 m, and 5.17 m for target sizes of 150 m, 500 m, 750 m, and 1 km, respectively.

We use the impactors SFDs of 500 m, 750 m and 1 km sized NEAs to compare with the observed craters on Bennu, Didymos and Ryugu by the OSIRIS‐REx, DART and Hayabusa2 missions. To do so, we convert impactor diameters to crater diameters using the scaling law for cohesive soil from \citet{Holsapple2007}:

\begin{equation}
d_{\mathrm{crat}}=K_{1}\left[ \left( \frac{gd_\mathrm{imp}}{2v^{2}}\right)\left(\frac{\rho_\mathrm{tgt}}{\rho_\mathrm{imp}}\right)^{\frac{2\nu}{\mu}}+\left(\frac{Y}{\rho_\mathrm{tgt}v^{2}} \right)^{\frac{2+\mu}{2}} \left(\frac{\rho_\mathrm{tgt}}{\rho_\mathrm{imp}}\right)^{\frac{\nu \left(2+\mu\right)}{\mu}}  \right]^{-\frac{\mu}{2+\mu}}d_\mathrm{imp},
\label{eq:Crater}
\end{equation}

with $\mu=0.41$, $\nu=0.4$, $K=1.03$, and we assume densities of $\rho_{\mathrm{imp}} = 2500 \,\mathrm{kg\,m^{-3}}$ and $\rho_{\mathrm{tgt}} = 1190 \,\mathrm{kg\,m^{-3}}$, $v=7.5 \mathrm{\,km\,s^{-1}}$ is the impact velocity \citep{campo2024recent}, $d_\mathrm{imp}$ is the impactor diameter in meters, and $Y$ and $g$ are the target strengths and surface gravities, respectively. 

\begin{figure*}
    \centering
    \includegraphics[width=18cm]{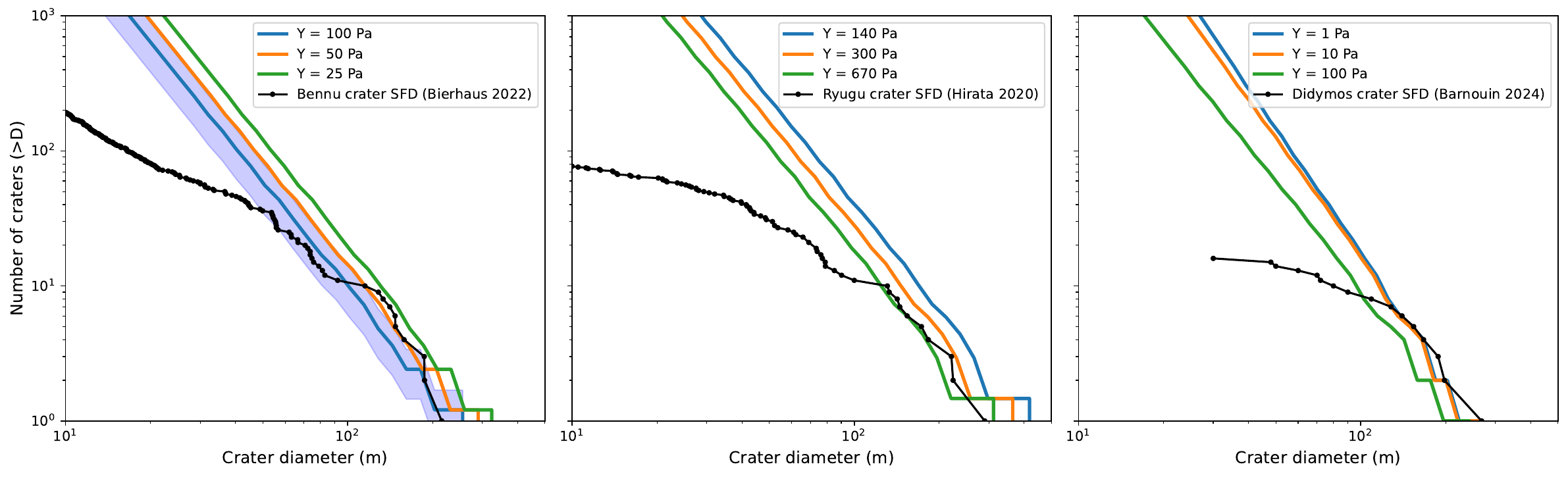}
    \caption{SFDs of craters produced on NEA targets with diameters of 500 m (left panel), 1 km (middle panel), and 750 m (right panel), corresponding to Bennu, Ryugu, and Didymos, respectively. Coloured solid curves show the median number of craters per Myr larger than a given diameter $D$ for different values of the target strength $Y$. Black curves represent the observed crater SFDs on Bennu \citep{bierhaus2022crater}, Ryugu \cite{hirata2020spatial} and Didymos \citep{barnouin2024}. Shaded area in left panel indicates the uncertainty in Bennu's lifetime as a NEA \citep{Ballouz2020}.}

    \label{fig:CraterSFD}
\end{figure*}

\subsection{Cratering on Bennu}

Bennu was the target of the OSIRIS-REx mission, which collected a sample from the asteroid and returned it to Earth on September 24, 2023. Multiple mineralogical analyses were then performed, accurately characterizing its mineralogical, chemical, and physical properties. The samples returned from asteroid Bennu confirm it as a highly aqueously altered, carbon‐rich body \citep{Lauretta2024}. The OSIRIS‐REx mission also mapped Bennu’s global chemistry, mineralogy, physical properties, and geology. Shape modeling from OSIRIS‐REx imagery—together with its high macroporosity and abundant surface boulders—confirms the predicted rubble‐pile structure \citep{Barnouin2019}. The mission’s global crater survey identified 1,560 craters ranging from under 1 m to over 200 m in diameter \citep{bierhaus2022crater}. \citet{bierhaus2022crater} showed that crater number densities drop sharply below diameters of $\sim$ 3 m—not due to observational bias or erasure processes, but likely because surface boulders ‘armour’ the regolith and inhibit small‐crater formation. This $\sim$ 3 m cutoff also constrains Bennu’s small‐crater retention age to between 1.6 and 2.2 Myr.

\cite{bierhaus2022crater} also reported a decrease in the density of large craters around a diameter of $\sim 100$ m and derived a crater-retention age of 10–65 Myr for the largest craters. The general shape of the SFD of craters observed on Bennu, with decreases near $\sim 3$ m and $\sim 100$ m, together with the fact that the overall crater SFD does not follow a single power law, suggests that the majority of Bennu’s crater population reflects two epochs: large craters that accumulated while Bennu resided in the MB and small craters produced during Bennu’s residence as an NEA. In this regard, our calculation of collisions on Bennu while it has been an NEA in a belt-crossing orbit could help constrain some presently undefined properties of this body, such as the target strength value as we will see below.

Bennu has been an NEA for $1.75 \pm 0.75$ Myr \citep{Ballouz2020}. Although its present orbit does not cross the MB, its semi-major axis places it within our Group~1 (G1; $a<2.06$ au), whose members frequently undergo temporary incursions into and out of the belt, as shown in Sect.~2. Objects in this group have a median fraction of orbital period spent inside the belt of $\Delta P_{\rm MB}/P \sim 0.55$, but with large object-to-object variability. Therefore, we computed the fraction $\Delta P_{\rm MB}/P$ specifically for Bennu. To do this, we generated 99 clones by randomly sampling Bennu's orbital elements within their observational uncertainties. The nominal orbit and its clones were then integrated for $3 \times 10^7$ years as massless particles under the gravitational influence of the Sun and the eight planets of the Solar System. For each particle and at each time step we computed $\Delta P_{\rm MB}/P$, and then we derived the median value of this quantity, obtaining $\Delta P_{\rm MB}/P = 0.36$. Adopting Bennu’s estimated NEA residence time of $1.75$ Myr, this corresponds to an effective MB exposure time of approximately $0.63$ Myr. Furthermore, because Bennu’s current orbit differs from that of Didymos, which was adopted as a reference for the collisional calculations performed in this work, we computed the intrinsic collision probability for Bennu following the same method and prescriptions of \cite{campo2024recent}, obtaining $P_i = 1.918 \times 10^{-17}\,\mathrm{km^{-2}\,yr^{-1}}$.

The median crater SFDs for a 500 m diameter asteroid representative of Bennu over an exposure time of 0.63 Myr are shown in the left panel of Fig.~\ref{fig:CraterSFD} for three target strengths, $Y=100, 50,$ and $25$ Pa. These strengths were selected based on the findings of \citet{perry2021impact}, who, using OSIRIS-REx observations of ejecta deposits surrounding a crater, determined that Bennu’s surface strength is likely lower than 100 Pa. We compare these results with the observed Bennu crater SFD from \citet{bierhaus2022crater}. For crater diameters $D \gtrsim 100$ m, the observations lie between the $Y=25$ Pa and $Y=50$ Pa curves. However, crater retention ages for features larger than 100 m on Bennu exceed 10 Myr \citep{bierhaus2022crater}, indicating that these craters most likely formed prior to Bennu’s residence as an NEA. We therefore consider $Y \approx 100$ Pa to be the most appropriate target strength for Bennu, as the observed crater distribution between 50 m and 100 m lies closest to the $Y=100$ Pa curve and shows a similar slope, and remains within the uncertainty associated with Bennu’s residence time as an NEA (blue shaded region in Fig.~\ref{fig:CraterSFD}). At smaller diameters, the model overpredicts crater counts, as expected given that our calculations do not include resurfacing or armoring processes that may erase small craters on rubble-pile asteroids.

\subsection{Cratering on Ryugu}

Asteroid Ryugu was the target of the Hayabusa2 mission, which returned surface samples from this NEA to Earth. Ryugu’s surface is covered by boulders from meter-scale up to $\approx 160$ m. The morphology and physical properties of these boulders indicate formation by catastrophic disruption and partial reaccumulation of a relatively homogeneous parent body, implying that Ryugu is a rubble-pile asteroid \citep{Watanabe2019}. Its bulk chemical composition is extremely primitive, close to solar and CI chondrites \citep{Nakamura2025}. Spacecraft spectral observations revealed a global abundance of hydrated minerals and clear evidence of aqueous alteration \citep{Kitazato2019}. 

The physical, mechanical, and thermal properties of the returned samples have been characterized in detail, enabling the construction of a model for the formation, evolution, and collisional disruption of Ryugu’s parent body. During the Hayabusa2 mission, a $\sim 30$ cm projectile (the Small Carry-on Impactor, SCI) was released at $\sim 2$ km s$^{-1}$, producing an artificial crater $>10$ m in diameter. Crater scaling indicates formation in the gravity regime; together with a low surface cohesion of $140$–$670$ Pa, this implies a surface age of $\sim 9$ Myr \citep{Arakawa2020}. \cite{hirata2020spatial} identified 77 impact craters larger than 20 m in diameter in images acquired by the Hayabusa2 spacecraft. Their analysis shows that the spatial distribution is heterogeneous: crater densities vary with longitude and latitude, and craters occur in clusters rather than being uniformly distributed.

As with Bennu, we computed the intrinsic collision probability between Ryugu and projectiles crossing its orbit obtaining $P_\mathrm{i}=1.41\times 10^{-17}\,\mathrm{km^{-2}\,yr^{-1}}$. The middle panel of Fig.~\ref{fig:CraterSFD} shows the derived crater SFD over an exposure time of 1 Myr on a 1 km, Ryugu-like target for three target cohesion values, $Y=140,\,300$ and $670$ Pa \citep{Arakawa2020}, together with the observed crater SFD on Ryugu's surface \citep{hirata2020spatial}. For $D\gtrsim100~\mathrm{m}$, the observed population lies between the curves for  $Y=300-670~\mathrm{Pa}$. Those large craters, however, were likely produced before Ryugu entered the NEA population \citep{takaki2022resurfacing}. At $D\lesssim100~\mathrm{m}$, all three $Y$ values overpredict the observed counts. Interpreting the model curves as crater production functions, the crater retention times can be estimated as $t(>D) = N_\mathrm{obs}(>D)/N_\mathrm{prod}(>D)$, which for Ryugu implies retention times $t \sim 10^4-10^5$ yr for $D<100$ m craters, consistent with rapid resurfacing \citep{takaki2022resurfacing}. 

\subsection{Cratering on Didymos}

(65803) Didymos was the target of the DART mission and is a binary asteroid system composed of a primary asteroid of the same name and its satellite Dimorphos. Images obtained by the DRACO camera during the final minutes of DART’s approach provided unprecedented coverage of both bodies, revealing surfaces covered in blocks, consistent with a rubble-pile structure.

\cite{barnouin2024} studied the surface geology of both components of the system. At higher latitudes, Didymos displays a rugged and undulating terrain, characterized by the presence of large boulders up to 160 m in length, degraded craters, and large-scale structures that may correspond to fossae or depressions. In the vicinity of the equator, a smooth region was identified that appears featureless at the resolution of the available images and lacks visible craters. This area may be composed of fine regolith or accumulations of blocks smaller than the resolution threshold, and its morphology suggests modification by surface mass-movement processes, likely associated with the asteroid’s rapid rotation, which promotes downslope transport of material toward the equator. In intermediate regions, evidence of surface transport processes is observed, such as tracks left by displaced blocks or small landslides \citep{bigot2024bearing}.

Overall, only 16 craters larger than 30 m have been identified, most of them located in higher-elevation regions. \cite{barnouin2024} found that the cohesive strength of the surface should be $< 2$ Pa, although at the center of Didymos the overburden stresses would exceed 100 Pa. They estimated a surface age of $\sim$12.5 Myr using values of $Y = 1$ Pa and $Y = 10$ Pa, which is slightly longer than the average lifetime in the NEA region. Therefore, the craters observed on Didymos should correspond not only to its time as a NEA but should also include craters formed when it was a MB asteroid. It is not possible to extrapolate with our method what occurred during the period when this asteroid belonged to the MB. However, when calculating the cratering over a time interval of 12.5 Myr, the number of craters predicted by our model is significantly larger than the number observed. According to our model, the calculated number of craters is consistent with the observations if the surface age is of order 1 Myr for strengths $Y$ of 1 Pa and 10 Pa, as shown in the right panel of Fig. \ref{fig:CraterSFD}. If this is the case, erosion processes on Didymos must be very efficient, and mass movements and even mass detachment associated with the asteroid’s rapid rotation would play a major role.

\cite{barnouin2024} also suggested that Didymos likely formed in near-Earth space, experiencing a major mass-shedding event about $0.09-0.3$ Myr ago. In this scenario, Didymos would have retained its largest craters, so the surface would not have been completely reset, although such an event could have modified the crater population. Furthermore, the presence of large surface blocks affects crater formation by inhibiting crater production \citep{bierhaus2022crater}, and the magnitude of this effect is still poorly understood. In the case of Didymos, the relationship between impactor size and crater size may therefore be altered by the abundance of blocks, so our estimates based on Eq. (\ref{eq:Crater}) may not fully capture the actual cratering process. Also, it is worth noting that the low number of identified craters on Didymos limits the level of analysis that can be performed and the conclusions that can be obtained using our method. Further observations and modeling efforts are required to accurately reproduce these effects. More detailed observations of the surface, such as those that will be provided by the HERA mission, will be crucial to better constrain both the cratering record and the surface processes operating on this asteroid.

\section{Conclusions}

In this study, we have combined $N$-body simulations with statistical collisional modeling to investigate the dynamical and collisional evolution of Near‐Earth Asteroids in main belt-crossing orbits (NEACs). These incursions represent a mean value of $60\%$ of their current orbital periods. By classifying these objects into three groups based on their initial semi‐major axis (G1: $a<2.06$ au; G2: $2.06<a<2.5$ au; G3: $a>2.5$ au), we draw the following main conclusions:

\begin{itemize}
  \item G3 asteroids show the shortest median lifetimes ($\sim1$ Myr), followed by G2 ($\sim3$ Myr) and G1 ($\sim20$ Myr), demonstrating that more compact orbits survive longer under planetary perturbations.
  \item G1 objects exhibit highly chaotic trajectories, frequently transitioning between belt‐crossing and belt‐confined orbits. Meanwhile, G2 and G3 predominantly act as belt-crossers and can transiently occupy orbits completely contained within the MB.
  \item The time NEACs spend in the MB, with respect to their orbital period, remains approximately constant during their lifetimes. Therefore, NEACs are susceptible to collisions with MB asteroids. 
  \item Collisional timescales for DART‐analog impacts are as short as $10^5$ yr for targets up to $\sim300$ m, increasing to $10^6$ yr for larger bodies. Catastrophic collisions on bodies larger than 100 m occur on timescales larger than $10^7$\,yr, and thus are negligible within the NEAC dynamical lifetimes.
  \item Collisional evolution of the NEAC population with MB asteroids over 1 Myr can erode between 0.1 \% and 1.4 \% of the size‐frequency distribution in the meter‐size range, depending on the $Q_\mathrm{D}^*$ prescription adopted.
  \item We derived the impactor SFDs for targets with diameters of 150 m, 500 m, 750 m, and 1 km, and used the 500 m, 750 m, and 1 km cases to construct crater SFDs for Bennu, Didymos, and Ryugu, respectively. For Bennu, the observations between 50 m and 100 m are best reproduced by a target strength $Y\sim100$ Pa, while the model overproduction of small craters is consistent with unmodeled resurfacing and armoring processes. For Ryugu, the modeled crater SFD implies short crater-retention times of order $10^{4}$--$10^{5}$ yr for $D<100$ m, in agreement with the resurfacing inferred by \cite{takaki2022resurfacing}. For Didymos, our results suggest that the observed crater population is consistent with effective surface ages shorter than the nominal NEA residence time, likely influenced by surface-modification processes such as mass movement and block-dominated terrains on a rapidly rotating rubble-pile asteroid.
  
\end{itemize}

The results indicate that the collisional activity of NEACs is too low to significantly alter the global properties of the population, such as its size distribution. Nevertheless, the study of collisional processes at the scale of small impacts remains important because of their implications for the asteroids’ physical, dynamical, and surface properties, and potential consequences for planetary defence strategies. These effects include changes in rotation rate, internal structure, crater formation, and dust ejection. Such processes will be examined in greater detail by the Hera mission, which will investigate the consequences of the DART impact on Dimorphos’s surface.

\begin{acknowledgements}
This research was performed using computational resources of Facultad de Ciencias Astronómicas y Geofísicas de La Plata (FCAGLP) and Instituto de Astrofísica de La Plata (IALP), and received partial funding from Universidad Nacional de La Plata (UNLP) under PID G172. Ricardo Gil-Hutton gratefully acknowledges financial support by CONICET through PIP 112-202001-01227.
\end{acknowledgements}

\bibliographystyle{aa} 
\bibliography{biblio} 

\end{document}